\documentclass{LT23auth}
\usepackage{graphicx}

\begin{document}

\begin{frontmatter}

\title{Spin-Ice and Other Frustrated Magnets on the Pyrochlore Lattice }

\author[address1]{ B Sriram Shastry\thanksref{thank1}}
\address[address1]{Indian Institute of Science, Bangalore 560042, INDIA }


\thanks[thank1]{ E-mail:bss@physics.iisc.ernet.in}

\begin{abstract}
The recent identification of the dysprosium titanate compound $Dy_2 Ti_2 O_7$ as a ``Spin-Ice'', i.e. the spin analog of regular entropic ice of Pauling, has created considerable excitement. The ability to manipulate spins using magnetic fields gives a unique advantage over regular ice in these systems, and has been used to study the recovery of entropy. Predicted magnetization plateaus have been observed, testing the underlying model consisting of a competition between short ranged super exchange, and long ranged dipolar interactions between spins. I discuss other compounds that are possibly spin ice like: $Ho_2 Ti_2 O_7$, and the two stannates $Ho_2 Sn_2 O_7$, $Dy_2 Sn_2 O_7$.
\end{abstract}

%
%
\begin{keyword}
Spin Ice  ; Zero Point Entropy  ; Ice Rule; DTO  
\end{keyword}
\end{frontmatter}

\section{ Introduction}
The recent realization that DTO ( $Dy_2 Ti_2 O_7$) is a spin analog of regular (  $I_h$) ice, has caused considerable excitement. Ice has fascinated several generations of physicists in view of its 
apparent violation of the Third Law of Thermodynamics, by virtue of having 
an entropic ground state. The calorimetry  experiment of Giauque and Stout \cite{gs} in 1936  was indeed one of the first triumphs of experimental low temperature physics, and  theory followed experiment 
rapidly.  The model of Pauling explained the origin of the entropy, as arising from the rearrangements of protons on the two possible locations on each $H-O-H$ bond, subject to the Bernal Fowler ice rule of two close protons 
and two far protons for each Oxygen on the wurtzite structure. DTO not only 
provides a spin realization of the two fold variable on each bond, it further 
gives one the handle of the Zeeman coupling between spins and the magnetic field, this energy   reduces the degeneracy, and the expected entropy recovery have stimulated considerable activity in the community. 
\begin{figure}[ht]
\begin{center}\leavevmode
\includegraphics[width=0.8\linewidth]{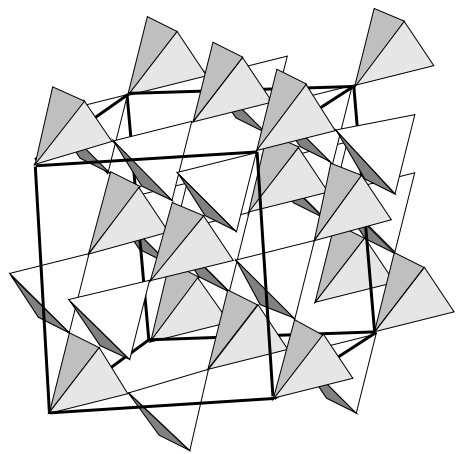}
\caption{ The pyrochlore lattice}
\label{fig0}
\end{center}
\end{figure}
In this 
talk I will give a quick review of the basic phenomena and discuss the modeling of the system. I will discuss the basic model that has been proposed by us, and tested against various experiments, most sensitively the magnetization plateaus seen recently. I will also mention some unresolved issues, mainly the predicted long ranged order in the true ground state that has  evaded observation by neutron scattering. After summarizing the situation of some other candidates for spin ice behaviour, most notably holmium titanate and the stannates, I briefly discuss the dynamical susceptibility that has been used to probe the nature of precursor spin liquid, i.e. the state that the spin ice melts into, it appears to be nontrivial in its correlations.  

\section{Basic Phenomenon and History}
 Early data \cite{gs} showed the surprising 
feature of entropy in the ground state, although crystallographic transitions
prevented the  cooling  of  $I_h$ ice in the wurtzite structure to very low temperatures. Bernal and Fowler pointed out the importance of minimizing dipolar energy in the proximity  of each Oxygen by formulating the famous ``ice rule'', i.e the rule that two protons are close by and two are further
from each Oxygen so that there is a six fold manifold of states that dominate 
the configuration space. Pauling  made the connection with experiments by estimating the global entropy arising from these local ice rules- the configurations on a single Oxygen influence the neighbours so there is a highly non trivial many body problem here. Pauling's estimate of the ground state entropy as $R \frac{1}{2} \log(\frac{3}{2})$ turns out
to be an inspired approximation in that that numerical computation of the entropy yield estimates that are  close. In the  world of models, this led to the celebrated two dimensional ice model, where the combinatorial 
problem was solved exactly by Lieb\cite{lieb} using {\em Bethe's Ansatz}. 
Anderson\cite{pwa} showed that the  identity  between the Spinel B site sublattice and the wurtzite structure leads to an interesting connection with 
the configurational entropy  in magnetite.

Harris {\em et al}\cite{harris} showed in 1997 that the pyrochlore compound 
HTO ( $ Ho_2 Ti_2 O_7$) could be considered as  ``spin ice''
by noting the absence of LRO in neutron scattering and from $\mu$SR
data. Moessner\cite{moess} emphasized  the fact that ferromagnetism plus a non collinear Ising type easy axis
arrangement can lead to frustration, which is commonly associated with 
{\em antiferromagnetism}. 
\begin{figure}[h]
\begin{center}\leavevmode
\includegraphics[width=0.8\linewidth]{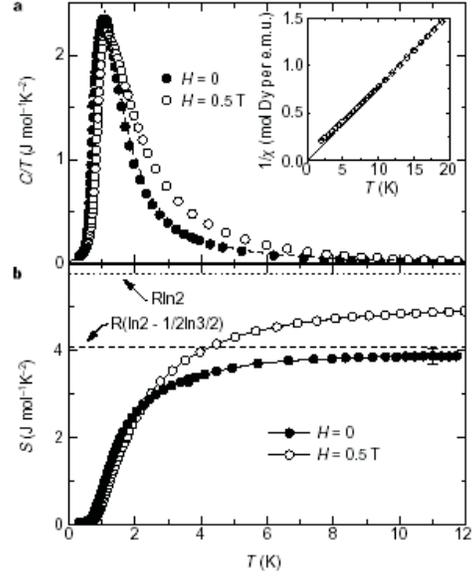}
\caption{The specific heat and integrated entropy from Ref(\cite{art}) }
\label{fig1}
\end{center}
\end{figure}

In a definitive experiment in 1999   Ramirez {\em et al}\cite{art} showed that the entropy of DTO 
has a shortfall from the expected $R \log(2)$ by an amount that is very close to the Pauling value, thus establishing it as the first clear spin ice
system. In Figure(2) we see that the entropy in zero field powder DTO saturates nicely, and a finite field, as small as .5T releases some of  the entropy. We return to discuss the details of ``entropy recovery'' later.

The currently popular  basic model to describe the spin ice system was formulated in Siddharthan {\em et al}\cite{rsidd}. It consists of local Ising variables pointing along or against the local easy axis found in each case by joining the sites to the center of either of the two tetrahedra it belongs to. These local moments interact via the long ranged magnetic dipole-dipole interaction in addition to the short ranged superexchange interaction $J$, which provides the single parameter of the theory. The values of the local moment are large for DTO (J=15/2 and $p= g J \mu_B$= 10 $\mu_B$) so the dipolar interaction energy is $\sim 2.3^0K$ for nearest neighbours.   The dipolar interaction is long ranged but  not 
a source of divergences  in this problem, since all the interesting physics is away from zero momentum (i.e. ferromagnetic region). An interesting point is that the ice rule configurations are increasingly important at lower temperatures, in Figure(3) we plot the fraction of tetrahedra in the ice rule configuration for three different values of $J$ taken from a crude monte carlo sampling. 
\begin{figure}[h]
\begin{center}\leavevmode
\includegraphics[width=0.8\linewidth]{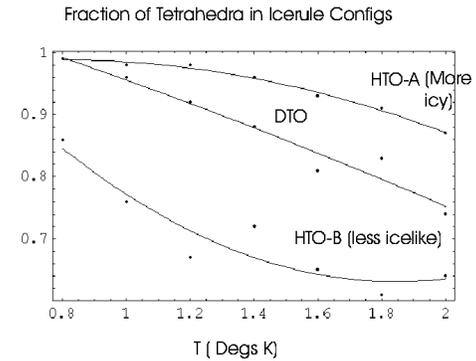}
\caption{The fraction of tetrahedra in ice ruled configurations for three values of superexchange $J=$ 1.9, 1.24, .52  for curves from bottom up.
  The middle curve corresponds to DTO, and the other to two possible parameters for HTO.}
\label{fig3}
\end{center}
\end{figure}
It is infact easy to see that the six configurations allowed by the ice rule possess a net magnetic moment that points along or antiparallel to the three
crystallographic axes, so at low temperature one expects a renormalization of the effective moment by a factor of $\frac{2}{3}$, and hence the uniform susceptibility by $\frac{4}{9}$. Data from Lawes and Ramirez\cite{lawes} on susceptibility is consistent with this expectation, as seen in Figure(4).
\begin{figure}[h]
\begin{center}\leavevmode
\includegraphics[width=0.8\linewidth]{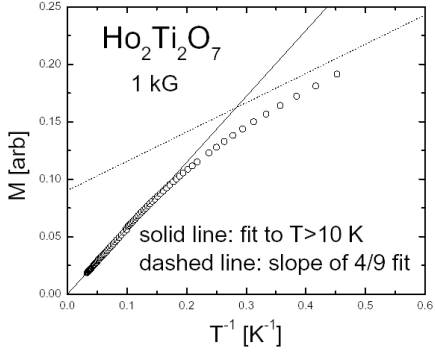}
\caption{The low temperature susceptibility of  HTO follows a Curie law but the Curie constant renormalizes to $\sim \frac{4}{9}$ below the Ice rule scale of temperature.}
\label{fig4}
\end{center}
\end{figure}

The agreement between experiments and theory on specific heat  is quite good
with the basic model\cite{rsidd,gingras}, 
and one can fine-tune the basic model by adding superexchange at further neighbour distance to improve  the agreement further\cite{rsiddthesis}. We note that the most stringent test is the behaviour of the magnetization versus magnetic field for different directions of the field.
The prediction of Siddharthan, Shastry and Ramirez\cite{ssr} of magnetization plateaux for fields along $<111>$ direction represent the best test, since the plateau at  $\sim 1 T$ corresponds to some tetrahedra breaking the ice rule configurations, and going into a 3-in 1-out configuration. Recent experiments by Matsuhira {\em et al}\cite{matsuhira1} bear out this prediction very well, as seen in Figure(5).
\begin{figure}[ht]
\begin{center}\leavevmode
\includegraphics[width=0.7\linewidth]{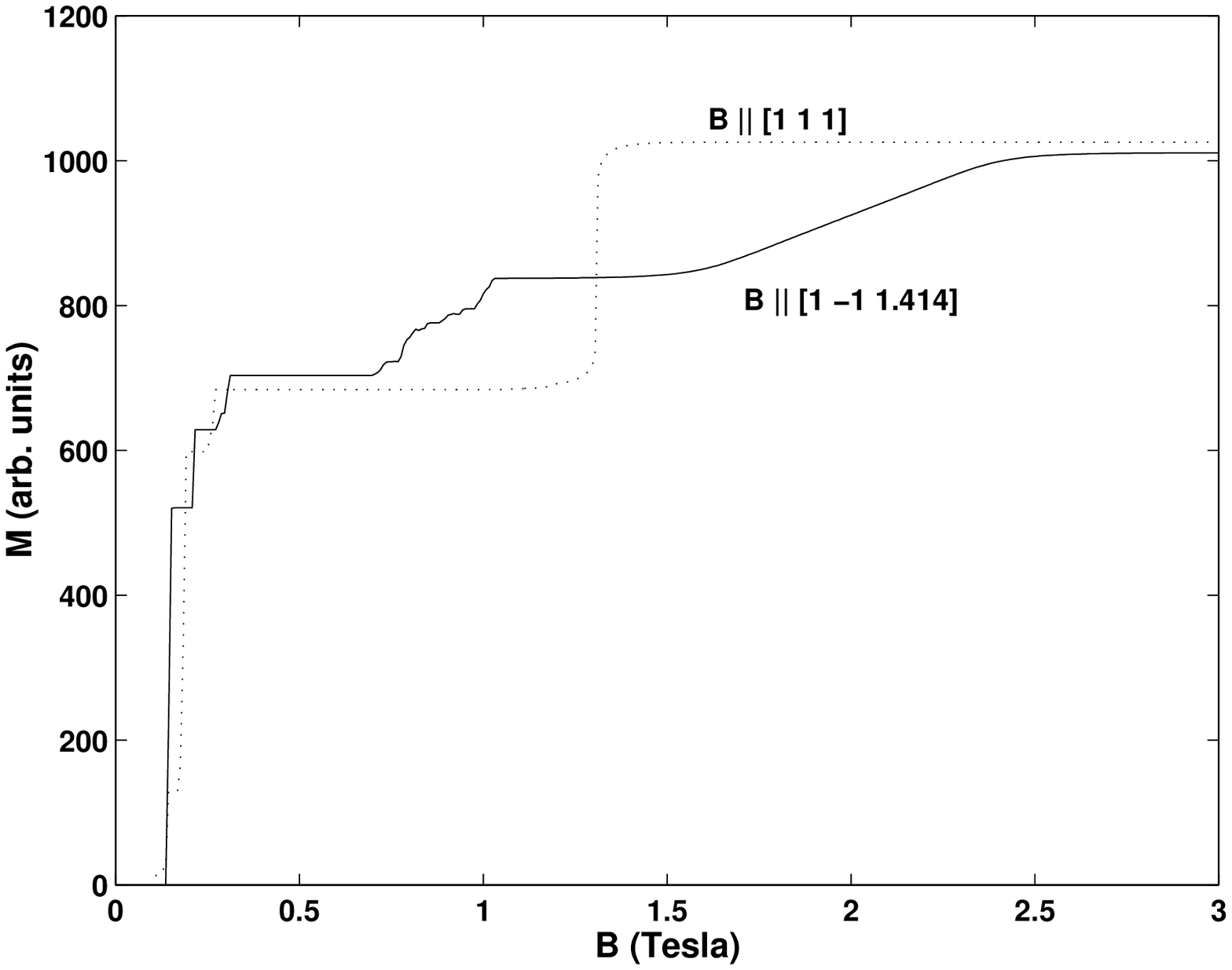}
\includegraphics[width=0.7\linewidth]{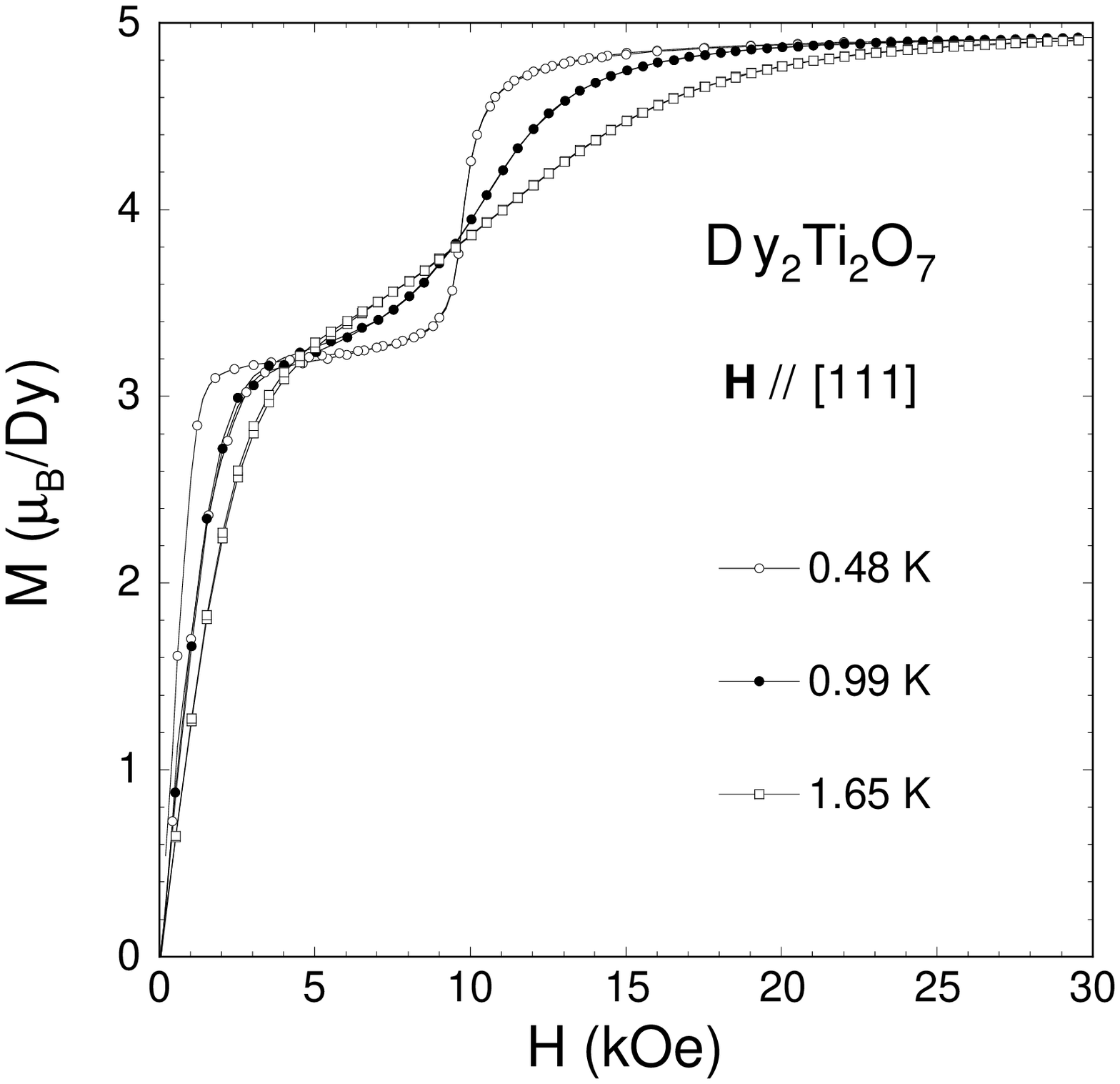}
\caption{Figure on top is the  theoretical prediction\cite{ssr}, the dotted
curve should be compared with 
the one below from experiment\cite{matsuhira1}. }
\label{fig5}
\end{center}
\end{figure}
The onset of the second plateau is sharp at low temperatures, and can be used to {\em  experimentally define the ice rule energy scale}. It should be regarded as the
fingerprint of the ice rule physics. This scale arises from the competition beween the (known) ferromagnetic dipolar energy and the (unknown) antiferromagnetic superexchange. In Figure(6) we show the range over which the ice rule scale 
changes by changing the exchange $J$, this can be used ( with linear extrapolation) to {\em deduce } $J$ if we know the location of the plateau.
\begin{figure}[ht]
\begin{center}\leavevmode
\includegraphics[{ width=0.6\linewidth,angle=90}]{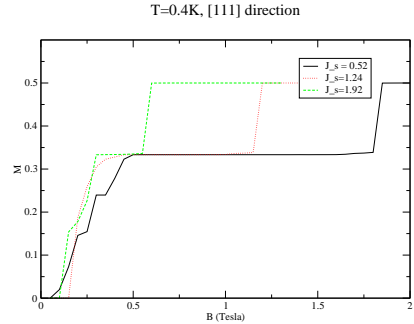}
\caption{Three values of $J$ and the resulting range of the plateau. the values of $J$ are $1.92,1.24,.52$ from left to right. }
\label{fig6}
\end{center}
\end{figure}
The values of the magnetic moment at the first and second plateaux are easy to understand, the saturation values are $\frac{1}{3}$ and $\frac{1}{2}$ of the maximum. 

The problem of entropy recovery in the presence of a uniform magnetic field was alluded to earlier. While the first experiment on this in Ref\cite{art} was on poweder samples, the monte carlo data \cite{ssr} already gave indications of considerable fine structure in the specific heat depending upon  the direction of the field relative to the   symmetry axes. For example a double peak structure was predicted for fields along $<1,-1,\sqrt{2}>$ direction which looks rather close to the double  peak seen 
very recently  in experiments\cite{matsuhira1}. 

We next mention  holmium titanate HTO, where $Ho^{3+}$ has an identical
magnetic moment as $Dy^{+3}$, and hence the same dipolar interaction, but possibly different exchange $J$. Initial experiments \cite{art}
showed a rising specific heat down at $\sim .5^0K$, at which point the
system fell out of equilibrium. One can interpret this in one of two possible ways. It is imaginable \cite{rsidd,rsiddthesis,ssr} that the system undergoes a transition to an long range ordered (LRO) state which is anyway predicted by theory (see  below), and that the slow dynamics hides this  transition. One has to then explain the difference between DTO and HTO as possibly arising from differences in the ice rule scale, or equivalently $J$, so that HTO would have a much lower ice rule temperature ( hence larger $|J| $) whereas DTO with a higher ice rule temperature would be stuck in a subset of configurations that would prevent it from undergoing a transition to a state with lower free energy.  The other view point\cite{bramwell} ascribes the difference in behaviour to the   significant hyperfine coupling constants in HTO, as known from early work of Blote {\em et al}\cite{blote}. Subtracting the nuclear component makes the data for HTO look similar to DTO, and the recent first principles calculations of the hyperfine constants of these two compounds \cite{ghosh} lends weight to this view point, as does the absence of LRO as indicated by neutron scattering \cite{bramwell,kadowaki}and other experiments at low temperatures $\sim 50^0 mK$.

Staying with the issue of LRO, we mention that even in the case of DTO one expects LRO at low enough temeratures. Siddharthan { \em et al} \cite{rsiddthesis,ssr} were the first to show, by explicit enumeration of ground states for finite clusters, that the model for DTO should have LRO of a certain type. This has been corraborated by Melko {\em et al}\cite{melko}, who use a loop algorithm to equilibriate the system rather than single spin flips, and find the same structure, with a transition at $\sim .2^0K$. In contrast, neutron scattering\cite{fennell} sees no signs of LRO down to  low temperatures $\sim 50^0 mK$! Therefore it seems to me that DTO and possibly HTO may yet have a surprise in store at low temperatures.

Turning to other possible spin ice compounds, the stannates $M_2 Sn_2 O_7$
with $M= Dy, Ho$ have been recently proposed \cite{matsuhira2}. These have small differences in lattice constants, and hence slightly different dipolar as well as superexchange interactions, and should be useful in fixing several details of the models.

Finally on the topic of spin ice, I would like to mention two very recent and nice experiments on the AC susceptibility of DTO. Schiffer {\em et al} \cite{schiffer} and Matsuhira {\em et al} \cite{matsuhira3} have found that the curie type divergence of the low field susceptibility ( as in Figure(4)) is infact cut off at a low temperature $\sim 15^0K$ that depends upon the 
frequency of the probe field, resulting in two maxima, one at a higher temperature $\sim 18^0K$ and the other at a lower temperature $\sim 1.5^0K$.
Each of these maxima is freqency dependent, and the from the Arrhenius 
dependence of these we can extract a characteristic  energy scale $\epsilon_c$ corresponding to 
these peaks. The higher one yields an energy scale $\sim 220^0K$, very nearly the crystal field splitting seen in \cite{rsidd}, and the lower one give 
$\sim 10^0K$, presumably a multiple of the ice rule energy scale. The real as well as imaginary parts of the susceptibility show the same features, and these seem to be consistent above $\sim 10^0K$ with a dynamical susceptibility $\frac{\chi_0}{(1-i \omega \tau)^\alpha}$, where $\chi_0 \sim \frac{1}{T}$ and
$\alpha $ departing from the Debye value of unity, and further being rather temperature dependent, with $\alpha \sim .544$ at $17^0K$ and $d \alpha/ dT \sim .06/ ^0K$.
 Thus the non Debye relaxation with temperature dependent $\alpha$, and the origin of the relaxation rate $\tau \sim \tau_0 \exp(\epsilon_c/k T)$ demand a fundamental understanding that is missing at the moment.

\section{}
\begin{ack}
It is a pleasure to acknowledge my collaborators:  Art Ramirez and Gavin Lawes  at Los Alamos National Lab, and Rahul Siddharthan at IISc/ENS Paris.
\end{ack}

%
%

\end{document}